\documentclass[12pt]{article}

\usepackage[totalheight = 23cm, totalwidth = 17cm]{geometry}
\usepackage{amssymb,amsmath,amsfonts,amsbsy,graphicx,xcolor,axodraw}
\usepackage{bm}

\newcommand{\mpl}{m_{\rm Pl}}
\newcommand{\fnl}{f_{\rm NL}}
\newcommand{\calA}{{\cal A}}
\newcommand{\calC}{{\cal C}}
\newcommand{\calD}{{\cal D}}
\newcommand{\calH}{{\cal H}}
\newcommand{\calI}{{\cal I}}
\newcommand{\calL}{{\cal L}}
\newcommand{\calN}{{\cal N}}
\newcommand{\calO}{{\cal O}}
\newcommand{\calP}{{\cal P}}
\newcommand{\calR}{{\cal R}}
\newcommand{\calW}{{\cal W}}

\newcommand{\lk}{\textbf{k}}

\begin{document}

\begin{titlepage}

\rightline{\footnotesize{APCTP-Pre2015-008}} \vspace{-0.2cm}

\begin{center}

\vskip 1.0 cm

\LARGE{\bf Correlation of isocurvature perturbation and non-Gaussianity}

\vskip 1.0cm

{\large
Jinn-Ouk Gong$^{a,b}$ \hspace{0.2cm} and \hspace{0.2cm} Godfrey Leung$^{a}$
}

\vskip 0.5cm

\small{\it
$^{a}$Asia Pacific Center for Theoretical Physics, Pohang 790-784, Korea
\\
$^{b}$Department of Physics, Postech, Pohang 790-784, Korea
}

\vskip 1.2cm

\end{center}

\begin{abstract}

We explore the correlations between primordial non-Gaussianity and isocurvature perturbation. We sketch the generic relation between the bispectrum of the curvature perturbation and the cross-correlation power spectrum in the presence of explicit couplings between the inflaton and another light field which gives rise to isocurvature perturbation. Using a concrete model of a Peccei-Quinn type field with generic gravitational couplings, we illustrate explicitly how the primordial bispectrum correlates with the cross-correlation power spectrum. Assuming the resulting $\fnl \sim \calO(1)$, we find that the form of the correlation depends mostly upon the inflation model but only weakly on the axion parameters, even though $\fnl$ itself does depend heavily on the axion parameters. 

\end{abstract}

\end{titlepage}

\newpage
\setcounter{page}{1}

\section{Introduction}

Inflation~\cite{inflation,Linde:1981mu} is widely considered as the best candidate to explain the origin of the primordial fluctuations in our universe, and has been well tested by cosmic microwave background (CMB) observations, most recently by the Planck mission~\cite{Ade:2015lrj}. The simplest scenario involves a single scalar field slowly rolling down a flat potential~\cite{Linde:1981mu}. However, when embedded in ultraviolet complete theories such as string theory~\cite{infreview}, it is not likely that such a simple description of single field inflation remains legitimate throughout the whole inflationary epoch, as in general additional degrees of freedom become dynamically relevant during inflation.

It is illustrative to consider the field trajectory in the field space to see how these additional degrees of freedom affect the inflationary predictions. While in single field inflation we have only a single direction along which the inflaton moves straightly, in multi-field space we have more than one direction and thus the trajectory is in general curved. As the component of field fluctuations along the trajectory is associated with the curvature perturbation, those orthogonal to the trajectory are responsible for the isocurvature perturbations~\cite{Gordon:2000hv}. What is important here is that since the field trajectory is in general curved, the fluctuation along the trajectory at a moment receives contributions from those orthogonal to the trajectory before. In other words, the curvature and isocurvature perturbations are correlated~\cite{Gordon:2000hv,correlation}.

Thus, on general ground the existence of isocurvature perturbations can be interpreted as a proof of additional degrees of freedom in the early universe. Even if the isocurvature perturbations themselves are too small to be detected, their correlation to the curvature perturbation can lead to distinctive observational signatures. It is thus useful and important to study how observables are correlated with and sourced by isocurvature perturbations on completely general ground. In this article, we concentrate on the correlations between primordial non-Gaussianity and isocurvature perturbations due to interactions at horizon-crossing, whereas previous studies focus on correlations due to non-linear evolution on super-horizon scales~\cite{other_ng-iso}. The primordial non-Gaussianity we are considering might not necessarily be that associated with the CMB scales, but also on smaller scales which are relevant for large scale structure for example. By studying the correlation structure, we can put further constraints on the model parameters compared to that coming from studying the observables individually. 

This article is organised as follows: in Section~\ref{sec:general_structure}, we discuss the general structure of three-point interaction terms between the inflaton and an isocurvature field, and how the resulting bispectrum $B_\calR$ correlates with the cross-correlation power spectrum $\calP_\calC$. We then illustrate this general feature using a concrete example involving a Peccei-Quinn (PQ) type field, which interacts with the inflaton via a gravitationally induced coupling in Section~\ref{sec:example}. We subsequently conclude in Section~\ref{sec:conc}. Detailed derivations of the results are given in the Appendix.

\section{General structure of correlations}
\label{sec:general_structure}

For simplicity, let us consider a generic two-field system, one being the inflaton $\phi$ and the other being a spectator field $\theta$ that survives after thermalisation and becomes responsible for the isocurvature perturbation $\calI$ at late times. We also assume the curvature perturbation $\calR$ is purely sourced by the inflaton fluctuations. In such a system, the three-point correlation function of the curvature perturbation $\calR$ may receive additional contributions from $\theta$~\cite{Bartolo:2001cw} via the quadratic interaction $H_2^I \sim \theta\phi$ which is responsible for the cross-correlation power spectrum $\calP_\calC$. The resulting bispectrum of the curvature perturbation $B_\calR$ is therefore correlated with $\calP_\calC$, giving additional constraints to the spectator field $\theta$.

\begin{figure}[h]
\begin{center}
 \begin{picture}(440,100)(0,0)
  \GCirc(40,60){20}{0.7}
  \Line(40,80)(40,100)
  \Line(10,30)(26,46)
  \Line(70,30)(54,46)
   \Text(85,60)[]{\large $=$}
  \Line(130,60)(130,100)
  \Line(100,30)(130,60) 
  \Line(160,30)(130,60)
   \Text(130,15)[]{$H_3^\text{(self1)} \sim \phi^3$}
  \Text(175,60)[]{\large $+$}
  \Line(220,60)(190,30)
  \Line(220,60)(250,30)
  \DashLine(220,60)(220,80){2}  \Line(220,80)(220,100)
   \Vertex(220,80){3}
   \Text(207,15)[]{1 $H_2^I$ and}
   \Text(222,0)[]{$H_3^\text{(cross1)} \sim g_1\theta\phi^2$}
  \Text(265,60)[]{\large $+$}
  \Line(310,60)(340,30)
  \DashLine(310,60)(310,80){2} \Line(310,80)(310,100)
   \Vertex(310,80){3}
  \DashLine(310,60)(295,45){2} \Line(280,30)(295,45)
   \Vertex(295,45){3}
   \Text(300,15)[]{2 $H_2^I$'s and}
   \Text(313,0)[]{$H_3^\text{(cross2)} \sim g_2\theta^2\phi$}
  \Text(355,60)[]{\large $+$}
  \DashLine(400,60)(400,80){2} \Line(400,80)(400,100)
   \Vertex(400,80){3}
  \DashLine(400,60)(385,45){2} \Line(385,45)(370,30)
   \Vertex(385,45){3}
  \DashLine(400,60)(415,45){2} \Line(415,45)(430,30)
   \Vertex(415,45){3}
   \Text(395,15)[]{3 $H_2^I$'s and}
   \Text(400,0)[]{$H_3^\text{(self2)} \sim g_3\theta^3$}
 \end{picture}
\end{center}
\vskip 0.3cm
\caption{Feynman diagrams for the contributions of $H_3$ given by \eqref{H3} to the inflaton three-point function. $H_3^\text{(self1)}$ is the intrinsic three-point inflaton self-interaction, which is of $\calO(\epsilon)$ and is usually slow-roll suppressed~\cite{Maldacena:2002vr}. $H_3^\text{(cross1)}$, $H_3^\text{(cross2)}$ and $H_3^\text{(self2)}$ contribute via interactions with the isocurvature field.}
\label{fig:interaction}
\end{figure}
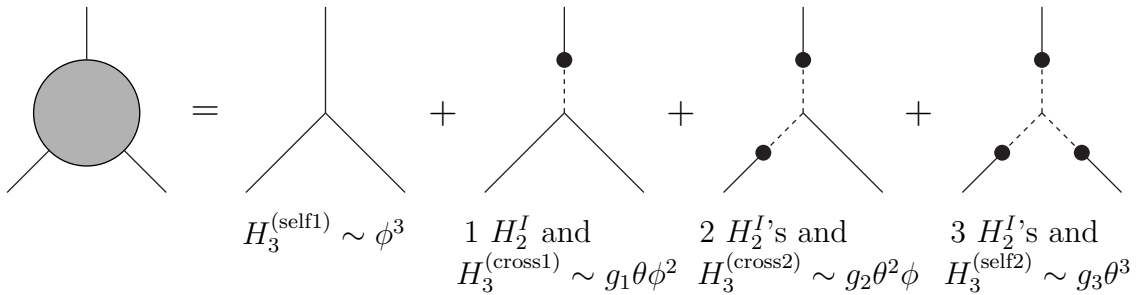

To see this explicitly, we decompose the cubic interaction Hamiltonian as 
\begin{equation}\label{H3}
H_3 = H_3^\text{(self1)} + H_3^\text{(cross1)} + H_3^\text{(cross2)} + H_3^\text{(self2)} \, ,
\end{equation}
where $H_3^\text{(self1)} \sim \phi^3$, $H_3^\text{(cross1)} \sim g_1\theta\phi^2$, $H_3^\text{(cross2)} \sim g_2\theta^2\phi$ and $H_3^\text{(self2)} \sim g_3\theta^3$ with $g_i$'s being the couplings. Note that these are schematic forms only, which could also include derivative interactions such as $(\nabla\theta)^2\phi$ and $\dot\theta^3$. In standard slow-roll inflation $H_3^\text{(self1)}$ usually contributes to a  slow-roll suppressed, negligible non-Gaussianity of $\calO(\epsilon)$~\cite{Maldacena:2002vr}. We therefore neglect $H_3^\text{(self1)}$ here. Through the transfer vertex $H_2^I$, each cubic interaction Hamiltonian term contributes to the primordial bispectrum $B_\calR$ as
\begin{equation}\label{generalcorr}
\begin{split}
H_3^\text{(cross1)} & \mapsto B_\calR \sim g_1\calP_\calR\calP_\calC \, ,
\\
H_3^\text{(cross2)} & \mapsto B_\calR \sim g_2\sqrt{\calP_\calR}\calP_\calC^2 \, ,
\\
H_3^\text{(self2)} & \mapsto B_\calR \sim g_3\calP_\calC^3 \, .
\end{split}
\end{equation}
Diagrammatic representation of each term is shown in Figure~\ref{fig:interaction}. However, the detail of each contribution, such as the momentum-dependent shape and amplitude, depends on the coupling and derivative structure of the corresponding cubic interaction Hamiltonian. In the following section using a concrete example we demonstrate that indeed the generic relation \eqref{generalcorr} is valid.

Before moving on to discuss a concrete example, let us comment on the relative size of the various bispectrum contributions above. Naively, one may think this somewhat suggests the above bispectrum contributions are small compared to that from the intrinsic inflaton self-interaction $H_3^\text{(self1)}$ since $H_3^\text{(self1)}\mapsto B_\calR \sim \calP_\calR^2$, if $\calP_\calR>\calP_\calC$. Besides, one might think there exists a hierarchy between the various contributions, with $B_\calR^\text{(cross1)}>B_\calR^\text{(cross2)}>B_\calR^\text{(self2)}$. However, this is not necessarily the case. In general, the resulting contributions are momentum and shape dependent. Depending on the forms of $H_3^\text{(cross1)}$, $H_3^\text{(cross2)}$ and $H_3^\text{(self2)}$, the various contributions might all have different momentum dependences and peak at different shapes. Furthermore, $\calP_\calC$ and $\calI$ also depend implicitly on the ratio of matter density of isocurvature modes to that of the total matter at late time, i.e. $\Omega_\theta/\Omega_m$, whereas the various bispectrum contributions above do not. If $\Omega_\theta/\Omega_m \ll 1$, it is possible to realise scenario where $\calP_\calC\ll \calP_\calR$ yet having $B_\calR^\text{(cross1)}$ $B_\calR^\text{(cross2)}$, $B_\calR^\text{(self2)}$ unsuppressed. 

\section{Axion with gravitationally induced interactions}
\label{sec:example}

Having discussed concisely the generic correlation structure between the bispectrum of the curvature perturbation $B_\calR$ and the cross-correlation power spectrum $\calP_\calC$ in a generic two-field system, we now explicitly illustrate this generic behaviour using a concrete model. The model Lagrangian in the matter sector is~\cite{Kadota:2014hpa}
\begin{equation}
\label{eq:the_model}
\calL = \sqrt{-g} \left[ -\frac{1}{2}(\partial_\mu\phi)^2 - |\partial_\mu\Phi|^2 - V_\text{inf}(\phi) - V_\text{axion}(\Phi) - V_\text{int}(\phi,\Phi) \right] \, ,
\end{equation}
where $\phi$ is the inflaton field and $V_\text{inf}(\phi)$ is its potential. We do not envisage any specific form of $V_\text{inf}(\phi)$, as long as it can successfully support an inflationary epoch. The complex field $\Phi$ is a PQ type field~\cite{PQ_field}, which can be decomposed into real radial and angular components, $r$ and $\theta$ respectively, as
\begin{equation}
\Phi = \frac{re^{i\theta}}{\sqrt{2}} \, ,
\end{equation}
The corresponding potential is the standard symmetry-breaking type, $V_\text{axion}(\Phi) = \lambda\left( |\Phi|^2 - f_a^2/2 \right)^2$, so that the radial field is minimised at $r_0 = f_a$ with $f_a$ being the symmetry breaking scale, which is usually taken to be smaller than $\mpl$. Assuming the radial field completely settles down at the minimum, we may identify the axion as the angular field $a \equiv f_a\theta$.

For concreteness, we consider a toy but generic dimension-5 interaction that can be gravitationally induced~\cite{Kamionkowski:1992mf},
\begin{equation}
\label{eq:V_interaction}
V_\text{int}(\phi,\Phi) = g\frac{\phi\Phi^4}{\mpl} + {\rm h.c.} \, .
\end{equation}
Additionally, overall we may add a cosmological constant to make the entire potential nearly vanishing at the global minimum. We shall associate the inflaton and axion fluctuations with the curvature and isocurvature perturbations respectively in the standard manner as~\cite{defn}
\begin{equation}
\label{eq:curv_iso_defn}
\calR \equiv -\frac{H}{\dot{\phi_0}}\phi \quad \text{and} \quad \calI \equiv \frac{2\theta}{\theta_0} \, .
\end{equation}
Here the subscript $0$ denotes the homogeneous background field values. We have implicitly assumed $\Omega_\theta/\Omega_m \sim 1$ as we are interested in the case where the isocurvature perturbations are large enough to be potentially observed in future experiments. 

Given the Lagrangian, we can straightly compute the correlation functions of the curvature and isocurvature perturbations using the in-in formalism~\cite{Maldacena:2002vr,in-in,Weinberg:2005vy}. The leading order contributions to the interaction Hamiltonian come from the gravitationally induced interaction term in the matter sector, which at quadratic and cubic orders are
\begin{align}
\label{eq:H2}
H_2^I = -a^3\alpha \int d^3x \phi\theta \quad \text{with} \quad \alpha & \equiv 2g\frac{r_0^4}{\mpl}\sin(4\theta_0) \, ,
\\
\label{eq:selfH3}
H_3^\text{(self2)} = a^3\beta \int d^3x \theta^3 \quad \text{with} \quad \beta & \equiv \frac{16}{3}g\frac{\phi_0}{\mpl}r_0^4\sin(4\theta_0) \, ,
\\
\label{eq:crossH3}
H_3^\text{(cross2)} = -a^3\gamma \int d^3x \phi\theta^2 \quad \text{with} \quad \gamma & \equiv 4g\frac{r_0^4}{\mpl}\cos(4\theta_0) \, .
\end{align}
The power spectra of the curvature and isocurvature perturbations, and their cross-correlation are 
\begin{align}
\label{eq:power_spectra_PR}
\calP_\calR(k) & = \left(\frac{H}{2\pi}\right)^2 \left(\frac{H}{\dot{\phi}_0}\right)^2 \left[ 2^{\nu_\phi-3/2} \frac{\Gamma(\nu_\phi)}{\Gamma(3/2)} \left(-k\tau\right)^{3/2-\nu_\phi} \right]^2 \, , 
\\
\label{eq:power_spectra_PI}
\calP_\calI(k) & = \left(\frac{H}{2\pi}\right)^2 \left(\frac{2}{\theta_0 r_0}\right)^2 \left[ 2^{\nu_\theta-3/2} \frac{\Gamma(\nu_\theta)}{\Gamma(3/2)} \left(-k\tau\right)^{3/2-\nu_\theta} \right]^2 \, ,
\\
\label{eq:power_spectra_PC}
\calP_\calC(k) & = -\frac{\pi\alpha}{2r_0H^2} \calA \sqrt{ \calP_\calR(k) \calP_\calI(k) } \, ,
\end{align}
where the indices of Hankel functions are defined as $\nu_\phi \equiv \sqrt{9/4 - m_\phi^2/H^2}$ with $m^2_\phi$ being the mass of $\phi$ and similarly for $\nu_\theta$. Here $\tau$ is the conformal time defined as ${\rm d}t=a{\rm d}\tau$, whereas $H$ and $\dot{\phi}_0$ are evaluated at horizon-crossing. Current observations suggest that the power spectra of isocurvature perturbations and the cross-correlation are much smaller than that of curvature perturbation on CMB scales, i.e. $\calP_\calR\gg \calP_\calC, \calP_\calI$, see the following. Here $\calA$ is an integral involving the Hankel functions, defined as
$
\calA \equiv \Re \left[ i\int_0^{\infty} dx/x H^{(2)}_{\nu_\phi}(x)H^{(2)}_{\nu_\theta}(x) \right]
$. 
If we naively perform this integral to leading order in $\nu_\phi \approx 3/2$ and $\nu_\theta \approx 3/2$ with a lower cutoff $x_c$, $\calA \sim \log{x_c}$ so it is logarithmically divergent. But taking into account the asymptotic behaviour of the curvature perturbation at later times, we may evaluate $\calA$ at an arbitrary value of $x_c$ as long as $\exp\left[-1/\calO(\epsilon)\right] < x_c < 1$~\cite{Gong:2001he}, giving $\calA \sim -0.45$. Details of the calculation are given in Appendix A.

Next we consider the primordial bispectrum of the curvature perturbation $B_\calR$ that arises from the cubic interaction Hamiltonian. There are two contributions: one from the axion cubic self-interaction $H_3^\text{(self2)} \sim \theta^3 $ and another from the cubic cross-interaction $H_3^\text{(cross2)} \sim \theta^2\phi$. We therefore expect the resulting bispectra scale as $\calP_\calC^3$ for the former interaction and $\calP_\calC^2\sqrt{\calP_\calR}$ for the latter one. After some  calculations similar as in~\cite{calculation}, we find the resulting bispectra of the curvature perturbation are of the form, to leading order in the limit $\nu_\phi\approx3/2$ and $\nu_\theta\approx3/2$,
\begin{align}
\label{eq:BcalR_self}
B_\calR^\text{(self2)}(k_1,k_2,k_3) & = \frac{k_1^3+k_2^3+k_3^3}{(k_1k_2k_3)^3} \calP_\calC^3  \frac{64\pi^{3}}{3\calA^3} \frac{\theta_0^3\beta}{H^4} \, ,
\\
\label{eq:BcalR_cross}
B_\calR^\text{(cross2)}(k_1,k_2,k_3) & = -\frac{k_1^3+k_2^3+k_3^3}{(k_1k_2k_3)^3} \calP_\calC^2\sqrt{\calP_\calR} \frac{16\pi^3}{3\calA^2} \frac{\theta_0^2\gamma}{H^3} \, .
\end{align}
Details of the calculation can be found in Appendix C. The above results of the two bispectrum contributions, $\calP_\calC$ and $\calP_\calR$ are evaluated on relevant scales when the gravitationally induced interaction \eqref{eq:V_interaction} becomes important, which do not necessarily correspond to CMB scales, but some smaller scales.

The amplitude of the bispectrum is usually measured in terms of the non-linear parameter $\fnl$~\cite{Komatsu:2001rj}. The current constraints on $\fnl$ by Planck are $\fnl^{\text{(local)}}=0.8\pm 5.0$, $\fnl^{\text{(eq)}}=-4\pm43$ and $\fnl^{\text{(ortho)}}=-26\pm21$ at 68\% confidence level~\cite{Ade:2015ava} for $k=0.05{\rm Mpc}$. Having in mind $\fnl$, we define a dimensionless shape function as
\begin{equation}
\label{eq:fnl_defn}
\fnl(k_1,k_2,k_3) \equiv \frac{10}{3} \frac{k_1k_2k_3}{k_1^3+k_2^3+k_3^3} \frac{(k_1k_2k_3)^2B_\calR}{(2\pi)^4\calP_\calR^2} \, .
\end{equation}
The corresponding shape functions for the bispectra $B_\calR^\text{(self2)}$ and $B_\calR^\text{(cross2)}$ are then
\begin{equation}
\label{fnls_1}
\fnl^\text{(self2)}(k_1,k_2,k_3) = \frac{40}{9\pi\calA^3} \frac{\calP_\calC^3}{\calP_\calR^2}  \frac{\theta_0^3\beta}{H^4}
\quad \text{and} \quad 
\fnl^\text{(cross2)}(k_1,k_2,k_3) = -\frac{10}{9\pi\calA^2} \frac{\calP_\calC^2}{\calP_\calR^{3/2}}  \frac{\theta_0^2\gamma}{H^3} \, ,
\end{equation}
or purely in terms of the model parameters as
\begin{equation}
\label{fnls}
\fnl^\text{(self2)} \sim  \left(\frac{\dot{\phi_0}}{H}\right) \left(\frac{\alpha}{r_0 H^2} \right)^3 \left(\frac{\beta}{r_0^3H^2} \right)
\quad \text{and} \quad 
\fnl^\text{(cross2)}\sim \left(\frac{\dot{\phi_0}}{H}\right) \left(\frac{\alpha}{r_0 H^2} \right)^2 \left(\frac{\gamma}{r_0^2H^2} \right)\, ,
\end{equation}
Again \eqref{fnls_1} might suggest $|\fnl^\text{(self2)}|<|\fnl^\text{(cross2)}|$ since $\calP_\calC<\calP_\calR$. This is not necessarily the case though as both of the non-linear parameters depend upon the coefficients $\beta$ and $\gamma$, which implicitly depend on the axion model parameters such as the misalignment angle $\theta_0$, as seen in \eqref{fnls}. It is possible to tune the axion model parameters such that $|f_{\rm NL}^{\rm (self2)}|>|f_{\rm NL}^{\rm (cross2)}|$ while having $\calP_\calC$ small.

In the following we give an estimate about the size of contributions to the non-linear parameter $\fnl$ on CMB scales in simple inflationary models. Assuming simple chaotic inflation with a quadratic potential, we have $\phi_0=15\mpl$ and the Hubble parameter at horizon-exit $H\approx 10^{-4}\mpl$. From \eqref{fnls} one then deduce 
\begin{align}
\label{fnls_example}
\fnl^\text{(self2)} \sim &  - \frac{1280}{27} g^4 \sin^4(4\theta_0) (\sqrt{2\epsilon}) \left(\frac{\phi_0}{\mpl} \right) \left(\frac{r_0}{\mpl} \right)^2\left(\frac{r_0}{H} \right)^8 \, , \nonumber \\
\fnl^\text{(cross2)}\sim & \,16 g^4 \sin^2(4\theta_0) \cos(4\theta_0) (\sqrt{2\epsilon})  \left(\frac{r_0}{\mpl} \right)^2\left(\frac{r_0}{H} \right)^6  \, ,
\end{align}
and $|\fnl^\text{(self2)}/\fnl^\text{(cross2)}| \sim g \tan(4\theta_0)\sin(4\theta_0) (\phi_0/\mpl)(r_0/H)^2$. Note that $r_0>H$ here since we are considering scenario where spontaneous symmetry breaking happens at higher energy scales than $H$. Therefore we expect $|\fnl^\text{(self2)}|>|\fnl^\text{(cross2)}|$ in this simple example unless the coupling is very small, i.e. $g\lesssim 10^{-3}$. In theory the value of $\fnl^\text{(self2)}$ could span a wide range depending on the combination of $r_0$ and $g$. For instance, assuming the symmetry breaking scale is around the GUT scale such that $r_0\sim 10^{-2}\mpl$, then we find $\fnl^\text{(self2)}\sim 1$ for $g\sim \calO(10^{-3})$. 

From \eqref{fnls_1}, we can also see that the resulting non-linear parameters are shape independent as opposed to non-canonical single field models such as K-inflation, where the non-Gaussianity induced due to non-linear interactions at horizon-crossing (or equivalently reduced sound speed $c_s$) typically peaks in the equilateral shape~\cite{nonG_cs}. This is because the forms of non-linear interactions are completely different between our case and the single field case. An another example where non-Gaussianity induced by non-linear interactions at horizon-crossing does not peak exactly in the equilateral shape is quasi-single field inflation~\cite{Chen:2009we}. In general models, however, the resulting non-linear parameters are not necessarily shape independent as we discussed in the last section. For our model, since $\fnl$ is shape independent, the tightest observational constraint comes from $\fnl^{\text{(local)}}$ above if the gravitationally induced interaction becomes relevant on CMB scales.

It is also convenient to express $\calP_\calC$ in terms of fractions of the isocurvature perturbation and cross-correlation, defined as\footnote{Note this definition of $\beta_\calI$ is slightly different from $\beta_\text{iso}$ used in~\cite{Ade:2015lrj}. They are related by $\beta_\calI = \beta_\text{iso}/\left(1-\beta_\text{iso}\right)$.}
\begin{equation}
\label{eq:def_fractions}
\beta_\calI \equiv \frac{\calP_\calI}{\calP_\calR} 
\quad \text{and} \quad 
\beta_\calC \equiv \frac{\calP_\calC}{\sqrt{\calP_\calR\calP_\calI}} \, .
\end{equation}
For generic cold dark matter isocurvature perturbation, at $k = 0.002 \text{Mpc}^{-1}$ the upper bound from Planck TT, TE, EE + lowP + WP is $\beta_\calI \lesssim 0.021$~\cite{Ade:2015lrj}. On the other hand,  the constraint on $\beta_\calC$ from Planck TT, TE, EE + lowP + WP is $-0.07\lesssim\beta_\calC\lesssim 0.21$~\cite{Ade:2015lrj}. Recent study has shown $|\beta_\calC|$ needs to be of $\calO(0.1)$ for forthcoming CMB experiments to be sensitive to the isocurvature cross-correlation~\cite{Kadota:2014hpa}. With the definitions \eqref{eq:def_fractions}, we can see how the observables are correlated in general
\begin{equation}
\fnl^{\text{(self2)}} \sim \beta_\calC^3\beta_\calI^{3/2}\calP_\calR^{5/2}
\quad \text{and} \quad
\fnl^{\text{(cross2)}} \sim \beta_\calC^2\beta_\calI \calP_\calR^{1/2} \, .
\end{equation}

\subsection{Consistency checks}
\label{sec:consistency}

There are implicit non-trivial constraints for the perturbation theory to remain valid, which also constrain the conditions for realising a large $\fnl$ given by \eqref{fnls}. In order to treat the transfer vertex $H_2^I$ as a small perturbation, the correction to the curvature perturbation power spectrum $\calP_\calR$ due to $H_2^I$ must be subleading compared to the leading piece \eqref{eq:power_spectra_PR}. Explicitly, the correction is given by
\begin{equation}\label{eq:deltaPR}
\Delta\calP_\calR(k) = \left( -\frac{\pi\alpha}{2r_0H^2} \right)^2 \calD \calP_\calR(k) \, ,
\end{equation}
where $\calD$ is an integral involving the Hankel functions defined as
\begin{equation}
\label{eq:calD}
\calD \equiv  \Re\left\{\int_0^\infty\frac{{\rm d}x_1}{x_1}\left[H_{\nu_\phi}^{(1)}(x_1) + H_{\nu_\phi}^{(2)}(x_1) \right] H_{\nu_\theta}^{(1)}(x_1) \int_{x_1}^{\infty} \frac{{\rm d}x_2}{x_2}  H_{\nu_\phi}^{(2)}(x_2) H_{\nu_\theta}^{(2)}(x_2)\right\} \sim \frac{(\log x_c)^2}{2!} \, ,
\end{equation}
with $x_c$ being the lower cutoff, so that approximately $\calD \sim \calA^2$. Thus for perturbative calculations to be valid, from \eqref{eq:deltaPR} we require 
\begin{equation}
\label{eq:condition_2point}
\left| \frac{\pi\alpha}{2r_0H^2} \right| \lesssim \calO(1) \, .
\end{equation}
Notice that this also means $\calP_\calC \lesssim \sqrt{\calP_\calR\calP_\calI}$. For detailed derivations, see Appendix B.

From \eqref{fnls}, the non-linear parameter $\fnl^{\text{(self2)}}$ given in terms of the model parameter scales as $\alpha^3\beta/(r_0^6H^8)$. Since the coefficients $\alpha$ and $\beta$ are in fact related by $\beta = -(4\phi_0/3)\alpha$ in our model, by using \eqref{eq:condition_2point} and taking $r_0=f_a$, we can see $\fnl^{\text{(self2)}}$ is as large as $\calO(1)$ only if
\begin{equation}
\label{eq:fnl_self_large_general}
\frac{\phi_0\mpl}{f_a^2} \gg 1  \, .
\end{equation}
This can be achieved by setting $\phi_0 \sim \calO(\mpl)$ and $f_a\ll \calO(\mpl)$. Similarly, the coupling $\gamma$ is related to the mass of the PQ field $m_\theta$. The massless axion limit, i.e. $m_\theta^2/H^2 \rightarrow 0$, thus corresponds to   
\begin{equation}
\label{eq:condition_m_theta}
\frac{\phi_0\gamma}{r_0^2H^2} \ll 1 \, .
\end{equation}
From \eqref{fnls}, the non-linear parameter $\fnl^{\text{(cross2)}}$ given in terms of the model parameter scales as $\alpha^2\gamma/(r_0^4H^6)$. As a result, we can see $\fnl^{\text{(cross2)}}\sim \calO(1)$ only if
\begin{equation}
\label{eq:fnl_cross_large_general}
\frac{\mpl}{\phi_0} \gg 1   \, ,
\end{equation}
which is possible only if $\phi_0$ is sub-Planckian. We can see the two conditions \eqref{eq:fnl_self_large_general} and \eqref{eq:fnl_cross_large_general} are mutually exclusive in general. We therefore conclude only one of them can be made large and the resulting non-linear parameter $\fnl$ either scales as $\calP_\calC^3$ or $\calP_\calC^2\sqrt{\calP_\calR}$. These results appear mainly due to the fact that the quadratic and cubic interaction Hamiltonians all arise from the same gravitationally induced interaction term \eqref{eq:V_interaction}.

Our results also apply to more general cases where the gravitationally induced interaction between the inflaton $\phi$ and the PQ field $\Phi$ is of the form
\begin{equation}
g \frac{\phi^m\Phi^n}{\mpl^{m+n-4}} \, .
\end{equation}
For $m\geq 2$, the cross-interaction $\phi^2\theta$ is also present, and contributes to $\fnl$ which is directly correlated with $\beta_\calC$,
\begin{equation}
\fnl^{\text{(cross1)}} \sim \beta_\calC \, .
\end{equation}
Whether or not this contribution can be comparable to $\fnl^{\text{(self2)}}$ and/or $\fnl^{\text{(cross2)}}$ depends upon $\phi_0$, which is model-dependent. The relative size between $\fnl^{\text{(cross1)}}$ and $\fnl^{\text{(self2)}}$ depends on the model of inflation with $\fnl^{\text{(cross1)}}/\fnl^{\text{(self2)}}\sim \mpl/\phi_0$, whereas for $\fnl^{\text{(cross2)}}$ the ratio go as $\sim\tan(n\theta_0)$.

\section{Conclusions}
\label{sec:conc}

In this article, 
we have studied the correlations between primordial non-Gaussianity and isocurvature perturbation due to horizon-crossing interactions on general ground. In the presence of explicit couplings between the inflaton and another light field that later produces isocurvature perturbation, the bispectrum of the curvature perturbation receives contributions from the conversion of isocurvature perturbation. These contributions can be written in terms of the cross-correlation power spectrum. Taking the gravitationally induced coupling of an axion-like field as an explicit example, we have shown that the primordial bispectrum correlates with the cross-correlation power spectrum as $B_\calR \sim \calP_\calC^3$ or $B_\calR \sim \calP_\calC^2\sqrt{\calP_\calR}$, depending on the inflation model. If we do observe primordial non-Gaussianity on smaller than CMB scales in future experiments, we might be able to say something about or even put constraints on inflation models where the inflaton interacts with axion-like fields in the early universe.

Besides the standard cold inflation scenario we have considered here, isocurvature perturbations can also arise naturally in warm inflation models via warm baryogenesis~\cite{warm_iso}. In such scenario, one should consider an inflaton-fluid system instead. We hope to investigate the possible correlations between the resulting isocurvature perturbation and primordial non-Gaussianity in such scenario in the future.

\subsection*{Acknowledgements}

We would like to thanks the referee for his/her valuable comments. We acknowledge the Max-Planck-Gesellschaft, the Korea Ministry of Education, Science and Technology, Gyeongsangbuk-Do and Pohang City for the support of the Independent Junior Research Group at the Asia Pacific Center for Theoretical Physics. 
This work is also supported in part by a Starting Grant through the Basic Science Research Program of the National Research Foundation of Korea (2013R1A1A1006701).

\section*{Appendix}

\subsection*{A. Computation of the two-point statistics}

Given the model Lagrangian \eqref{eq:the_model}, we can work out the corresponding Hamiltonian at appropriate order. At quadratic order, assuming slow-roll and the off-diagonal term of the mass matrix is small, we can take the free field Hamiltonian density, $\calH_{\rm free}$ , and the interaction Hamiltonian density, $\calH^I_2$ ,as
\begin{align}
\label{eq:free_int_Hamiltonian}
 \calH_{\rm free} = &  \frac{a^3}{2}\dot{\phi}^2 + \frac{a}{2} (\nabla\phi)^2 + \frac{a^3r_0^2}{2}\dot{\theta}^2 + \frac{a r_0^2}{2}(\nabla\theta)^2 + a^3\delta V_{\rm inf} - 8a^3 g \left(\frac{\phi_0r_0^4}{\mpl}\right)\cos(4\theta_0)\theta^2 \,, \nonumber \\
\calH^I_2 = &  -2a^3 g\frac{r_0^4}{\mpl} \sin(4\theta_0) \phi\theta =  -a^3 \alpha \phi\theta \,,
\end{align}
where $\alpha$ is defined as in \eqref{eq:H2}. The equations of motion of the free fields $\phi$ and $\theta$ follow from $\calH_{\rm free}$.

To compute the two-point correlation functions, we first promote the fields to operators and decompose them into annihilation and creation operators in Fourier space 
\begin{align}
\label{eq:field_perturbation_fourier_operator}
 \hat{\phi}_\lk = & \hat{a}_\lk u_\lk +  \hat{a}_{-\lk}^{\dagger} u^*_\lk  \,, \nonumber \\
 \hat{\theta}_\lk = & \hat{b}_\lk v_\lk +  \hat{b}_{-\lk}^{\dagger} v^*_\lk \,,
\end{align}
where $ \hat{a}_\lk$ and $ \hat{b}_\lk$ satisfying the commutation relations 
\begin{align}
\label{eq:operator_commutation}
\left[\hat{a}_\lk, \hat{a}_{\lk'}^{\dagger} \right] = \left[\hat{b}_\lk, \hat{b}_{\lk'}^{\dagger} \right] = (2\pi)^3\delta^{(3)}(\lk-\lk') \,,
\end{align}
and zero otherwise. The mode functions $u_\lk$ and $v_\lk$ satisfy the following equations 
\begin{align}
\label{eq:EOM_field_perturbation}
 u_\lk'' - \frac{2}{\tau} u_{\lk}' + \left(k^2 + \frac{m_\phi^2}{H^2\tau^2}\right) u_\lk =&  0 \,, \nonumber  \\
 v_\lk'' - \frac{2}{\tau} v_{\lk}' + \left(k^2 + \frac{m_\theta^2}{H^2\tau^2}\right) v_\lk =&  0 \,,
\end{align}
where 
\begin{equation}
\label{eq:mass_inflaton_axion}
m_\phi^2 \equiv \frac{\partial^2 V_{\rm inf}}{\partial\phi^2} \,\,\, {\rm and} \,\,\, m_\theta^2 \equiv -8g\frac{\phi_0}{\mpl}r_0^2 \cos(4\theta_0) \,.
\end{equation}

Assuming the last terms in \eqref{eq:EOM_field_perturbation} are small such that both fields behave as massless fields up to slow-roll approximation, we find the solutions of the mode functions 
\begin{align}
\label{eq:mode_func_soln}
u_\lk(\tau) =& -i \exp\left[i\left(\nu_\phi + \frac{1}{2} \right)\frac{\pi}{2}\right] \frac{\sqrt{\pi}}{2}H(-\tau)^{3/2}H^{(1)}_{\nu_\phi} (-k\tau) \\
v_\lk(\tau) =& -\frac{i}{r_0} \exp\left[i\left(\nu_\theta + \frac{1}{2} \right)\frac{\pi}{2}\right]\frac{\sqrt{\pi}}{2}H(-\tau)^{3/2}H^{(1)}_{\nu_\theta} (-k\tau) 
\end{align}
with $\nu_\phi \equiv \sqrt{9/4 - m_\phi^2/H^2}$ and similarly for $\nu_\theta$, for $0\leq \nu_j\leq 3/2$. Here $H^{(1)}_\nu$ is the Hankel function of the first kind. 

The correlation functions can then be computed using the in-in formalism~\cite{in-in}. Using the commutator form, the expectation values of some operators $\hat{O}(t)$ at the time $t$ is given by~\cite{Weinberg:2005vy}
\begin{align}
\label{eq:in_in_correlation_commutator}
\left\langle \hat{O} (t) \right\rangle = & \left\langle 0\bigg| \hat{O} (t) \bigg| 0 \right\rangle + i \int_{t*}^t {\rm d}t_1 \left\langle 0\bigg| \left[H_I(t_1), \hat{O} (t)\right] \bigg| 0 \right\rangle \nonumber \\
& +  i^2 \int_{t*}^t {\rm d}t_1 \int_{t*}^{t_1}{\rm d}t_2  \left\langle 0\bigg| \left[ H_{I}(t_1), \left[H_I(t_2), \hat{O} (t)\right] \right] \bigg| 0 \right\rangle + O(H_{I}^{3})
\end{align}
where $\left.|0\right\rangle$ is the interaction vacuum at some initial time $t_*$. The resulting power spectra \eqref{eq:power_spectra_PR}-\eqref{eq:power_spectra_PC}
can be found by inserting the interaction \eqref{eq:free_int_Hamiltonian} into \eqref{eq:in_in_correlation_commutator} for $\hat{O}=\left\{\hat{\phi}_{\lk_1}\hat{\phi}_{\lk_2}, \hat{\theta}_{\lk_1}\hat{\theta}_{\lk_2}, \hat{\phi}_{\lk_1}\hat{\theta}_{\lk_2}\right\}$. For instance, derivations of $\calP_\calC$ and $\calA$ can be found in~\cite{Kadota:2014hpa}. 

\subsection*{B. Computation of the correction to the leading order $\calP_\calR$}

In section~\ref{sec:consistency}, we argue that the condition \eqref{eq:condition_2point} have to be satisfied for consistency in order to treat $\calH^I_2$ as a small perturbation compared to $\calH_{\rm free}$. As stated in the section, this comes from the fact that the non-zero contribution to the inflaton power spectrum, or in other words, $\calP_\calR$, coming from $\calH^I_{2}$ should be subleading compared to the free-field contribution \eqref{eq:power_spectra_PR}. 

To see, we compute the correction to $\calP_\calR$ due to $\calH^I_{(2)}$ using the commutator form of the in-in formalism \eqref{eq:in_in_correlation_commutator} in a similar fashion as in~\cite{calculation}. Labeling this correction by $\Delta\calP_\calR$, this is of order $\calO[(\calH^{I}_{2})^2]$ and is given by

\begin{align}
\label{eq:inflaton_power_spectrum_correction}
\Delta\calP_\calR(k)  = &\int_{t_*}^t {\rm d}t_1\int_{t_*}^t {\rm d}t_2\left\langle 0 \left | H^I_{2}(t_1)\hat{\phi}_{\lk_1}(t)\hat{\phi}_{\lk_2}(t)H^I_{2}(t_2)\right |0\right\rangle \nonumber \\ 
& -2\Re\left[ \int_{t_*}^t {\rm d}t_1\int_{t_*}^{t_1} {\rm d}t_2\left\langle 0 \left |\hat{\phi}_{\lk_1}(t)\hat{\phi}_{\lk_2}(t)H^I_{2}(t_1
)H^I_{2}(t_2)\right |0\right\rangle \right] \, .
\end{align}
Assuming inflationary background such that $a \sim (-1/H\tau)$ and $H\approx {\rm const.}$ and taking the large wavelength limit after horizon-exit, we finally arrive \eqref{eq:deltaPR}
\begin{equation}
\label{eq:inflaton_power_spectrum_correction_final}
\Delta\calP_\calR(k)  = \left|u_\lk(0)\right|^2\frac{k^3}{2\pi^2} \left(\frac{\pi\alpha}{2r_0H^2} \right)^2  \calD =  \left( \frac{\pi\alpha}{2r_0H^2} \right)^2 \calD \calP_\calR(k) \, , \\
\end{equation}
where $\calD$ is an integral involving the Hankel functions defined as
\begin{equation}
\label{eq:calD}
\calD \equiv  \Re\left\{\int_0^\infty\frac{{\rm d}x_1}{x_1}\left[H_{\nu_\phi}^{(1)}(x_1) + H_{\nu_\phi}^{(2)}(x_1) \right] H_{\nu_\theta}^{(1)}(x_1) \int_{x_1}^{\infty} \frac{{\rm d}x_2}{x_2}  H_{\nu_\phi}^{(2)}(x_2) H_{\nu_\theta}^{(2)}(x_2)\right\} \, ,
\end{equation}
In the massless limit where $\nu_\phi, \nu_\theta \approx 3/2$, $\calD$ reduces to
\begin{equation}
\label{eq:inflaton_power_spectrum_correction_integral_massless}
\calD = \left(\frac{4}{\pi^2}\right) \Re\left\{\int_0^\infty \frac{{\rm d}x_1}{x_1^{4}} \left[ (1-i x_1)e^{i x_1} - {\rm c.c.}\right] (1-ix_1)e^{ix_1}  \int_{x_1}^{\infty} \frac{{\rm d}x_2}{x_2^{4}} (1+ix_2)^2e^{-2ix_2} \right\}  \,,
\end{equation}
and is logarithmically divergent in the IR limit. Here ${\rm c.c.}$ stands for complex conjugate. To regularsie this, we introduce a lower cut-off $x_c$ in the above integral. Then we can see 
$\calD$ approximately scale as
\begin{equation}
\label{eq:inflaton_power_spectrum_correction_integral_massless_IR}
\calD\sim \frac{(\log x_c)^2}{2!} \sim \calA^2 \,.
\end{equation}
For the correction to be small such that $\Delta\calP_\calR(k) / \calP_\calR(k) \ll \calO(1)$, we therefore generically require $|\pi\alpha/2r_0 H^2| \ll \calO(1)$, i.e. the condition \eqref{eq:condition_2point}.

\subsection*{C. Computation of the bispectrum}

The resulting primordial bispectrum can be computed in a similar fashion as the two-point functions. Assuming slow-roll, at cubic order, the corresponding interaction Hamiltonian $H_3$ dominated by the contribution from the gravitationally induced interaction \eqref{eq:V_interaction} and is given by 
\begin{align}
\label{eq:action_cubic_interaction}
H_3 = & \int d^3x \calH_3 = a^3\frac{g r_0^4}{\mpl} (2\pi)^{(3)} \delta^{(3)}(\lk_1 + \lk_2 + \lk_3)
\int \frac{d^3k_1}{(2\pi)^3}\int \frac{d^3k_2}{(2\pi)^3} \int \frac{d^3k_3}{(2\pi)^3}  \nonumber \\
\times & \left[ \frac{16}{3} \phi_0 \cos(4\theta_0) \hat{\theta}_{\lk_1}\hat{\theta}_{\lk_2}\hat{\theta}_{\lk_3} - \frac43 \sin(4\theta_0) \hat{\phi}_{\lk_1}\hat{\theta}_{\lk_2}\hat{\theta}_{\lk_3}  + {\rm perm.} \right]  \,, \nonumber \\
\end{align}
where ${\rm perm.}$ denotes permutation over momenta $\lk_j$ of the second term in the square bracket. Splitting $H_3$ in terms of the form of the interaction, we therefore arrive at \eqref{eq:selfH3} and \eqref{eq:crossH3}. 

\subsubsection*{C1. Contribution from $H_3^\text{(self2)}$}

Expressed in the nested commutator form of the in-in formalism, the leading contribution to the fully connected inflaton three-point function $\left\langle\phi^3 \right\rangle$ from the axion self-interaction Hamiltonian $H_3^\text{(self2)}$ is of cubic order in $H_2^I$ and is given by
\begin{equation}
\label{eq:3point_fun_inflaton_cross}
\left\langle \phi^3 \right\rangle^\text{(self2)}  = \int_{t_0}^{t} {\rm d}t_1 \int_{t_0}^{t_1} {\rm d}t_2 \int_{t_0}^{t_2} {\rm d}t_3\int_{t_0}^{t_3} {\rm d}t_4  \left\langle \left[H^I(t_4),\left[H^I(t_3),\left[H^I(t_2),\left[H^I(t_1),\hat{\phi}_I(t)^3 \right]\right]\right]\right]\right\rangle \,,
\end{equation}
where $H^I$ corresponds to an interaction Hamiltonian, with one being $H_3^\text{(self2)}$ and the others being $H_2^I$. The overall contribution is therefore~\cite{calculation}  
\begin{align}
\label{eq:3point_fun_inflaton}
\left\langle \phi^3 \right\rangle^\text{(self2)} = & 12\alpha^3\beta \left[u_{\lk_1}(0)u_{\lk_2}(0)u_{\lk_3}(0)\right] \Re \left[ \int_{-\infty}^0 d\tau_1 \int_{-\infty}^{\tau_1} d\tau_2 \int_{-\infty}^{\tau_2} d\tau_3\int_{-\infty}^{\tau_3} d\tau_4 \prod_{i=1}^4 a^{4}(\tau_i)\right. \nonumber \\
&\times \left.\left(A + B + C\right)\right] (2\pi)^3\delta^{(3)}(\sum_i \lk_i) + 5 \, {\rm perm.}  \, ,
\end{align}
where the terms $A$, $B$ and $C$ are
\begin{align}
\label{eq:3point_fun_inflaton_A_term}
A =&-[u_{\lk_1}(\tau_1) - {\rm c.c.}][v_{\lk_1}(\tau_1)v_{\lk_1}^*(\tau_2) - {\rm c.c.}][v_{\lk_3}(\tau_2)v_{\lk_3}^*(\tau_4)u_{\lk_3}(\tau_4) - {\rm c.c.}]  
v_{\lk_2}(\tau_2)v_{\lk_2}^*(\tau_3)u_{\lk_2}^*(\tau_3) \, , \\ 
\label{eq:3point_fun_inflaton_B_term}
B =&-[u_{\lk_1}(\tau_1) - {\rm c.c.}][u_{\lk_2}(\tau_2) - {\rm c.c.}][v_{\lk_1}^*(\tau_1)v_{\lk_2}^*(\tau_2)v_{\lk_1}(\tau_3)v_{\lk_2}(\tau_3) - {\rm c.c.}]
v_{\lk_3}(\tau_3)v_{\lk_3}^*(\tau_4)u_{\lk_3}^*(\tau_4) \, , \\ 
\label{eq:3point_fun_inflaton_C_term}
C =&[u_{\lk_1}(\tau_1) - {\rm c.c.}][u_{\lk_2}(\tau_2) - {\rm c.c.}][u_{\lk_3}(\tau_3) - {\rm c.c.}] 
v_{\lk_1}^*(\tau_1)v_{\lk_2}^*(\tau_2)v_{\lk_3}^*(\tau_3)v_{\lk_1}(\tau_4)v_{\lk_2}(\tau_4)v_{\lk_3}(\tau_4) \, , 
\end{align} 
corresponding to replacing $H^I(t_2)$, $H^I(t_3)$ or $H^I(t_4)$ with the axion self-interaction Hamiltonian $H_3^\text{(self2)}$ respectively. Note that replacing $H^I(t_1)$ with $H_3^\text{(self2)}$ gives zero contribution since $H_3^\text{(self2)}\propto \hat{\theta}_I^3$, which does not contract with any of the external legs $\hat{\phi}_I(t)$. Here we also have used the fact that $u_\lk(0)$ is real. Plugging in the solutions for the mode functions $u$ and $v$, \eqref{eq:3point_fun_inflaton_A_term}-\eqref{eq:3point_fun_inflaton_C_term} written in terms of the Hankel functions are
\begin{align}
\label{eq:3point_fun_inflaton_3_terms_Hankel}
A =& - \calW \left(-\tau_1\right)^3\left(-\tau_2\right)^{9/2}\left(-\tau_3\right)^3\left(-\tau_4\right)^3 \Re\left[H_{\nu_\phi}^{(1)}(-k_1\tau_1)\right] \Im \left[H_{\nu_\theta}^{(1)}(-k_1\tau_1)H_{\nu_\theta}^{(2)}(-k_1\tau_2)\right] \nonumber \\ 
\times &
\Re\left[H_{\nu_\theta}^{(1)}(-k_3\tau_2)H_{\nu_\theta}^{(2)}(-k_3\tau_4)H_{\nu_\phi}^{(2)}(-k_3\tau_4)\right]  
H_{\nu_\theta}^{(1)}(-k_2\tau_2)H_{\nu_\theta}^{(2)}(-k_2\tau_3)H_{\nu_\phi}^{(2)}(-k_2\tau_3) \, , \\ 
B =& \calW\left(-\tau_1\right)^3\left(-\tau_2\right)^{3}\left(-\tau_3\right)^{9/2}\left(-\tau_4\right)^3\Re\left[H_{\nu_\phi}^{(1)}(-k_1\tau_1)\right] \Re\left[H_{\nu_\phi}^{(1)}(-k_2\tau_2)\right] \nonumber \\ 
 \times & 
\Im \left[H_{\nu_\theta}^{(2)}(-k_1\tau_1)H_{\nu_\theta}^{(2)}(-k_2\tau_2)H_{\nu_\theta}^{(1)}(-k_1\tau_3)H_{\nu_\theta}^{(1)}(-k_2\tau_3)\right]  
H_{\nu_\theta}^{(1)}(-k_3\tau_3)H_{\nu_\theta}^{(2)}(-k_3\tau_4)H_{\nu_\phi}^{(2)}(-k_3\tau_4) \, ,
\\ 
C =& -i\calW\left(-\tau_1\right)^3\left(-\tau_2\right)^{3}\left(-\tau_3\right)^{3}\left(-\tau_4\right)^{9/2} \Re\left[H_{\nu_\phi}^{(1)}(-k_1\tau_1)\right] \Re\left[H_{\nu_\phi}^{(1)}(-k_2\tau_2)\right]\Re\left[H_{\nu_\phi}^{(1)}(-k_3\tau_3)\right] \nonumber \\ 
\times  & 
H_{\nu_\theta}^{(2)}(-k_1\tau_1)H_{\nu_\theta}^{(2)}(-k_2\tau_2)H_{\nu_\theta}^{(2)}(-k_3\tau_3)
H_{\nu_\theta}^{(1)}(-k_1\tau_4)H_{\nu_\theta}^{(1)}(-k_2\tau_4)H_{\nu_\theta}^{(1)}(-k_3\tau_4) \, ,
\end{align}
where we have defined $\calW \equiv 8/r_0^6(H\sqrt{\pi}/2)^9$. Assuming $\calR$ is sourced purely by the inflaton fluctuations, the resulting curvature perturbation bispectrum is simply 
\begin{equation}
\label{eq:3point_fun_R}
\left\langle \calR(\lk_1)\calR(\lk_2)\calR(\lk_3) \right\rangle = -\left(\frac{H}{\dot{\phi}_0}\right)^3 \left\langle\phi(\lk_1)\phi(\lk_2)\phi(\lk_3) \right\rangle \, ,
\end{equation}
where the corresponding bispectrum $B_\calR^\text{(self2)}(k_1,k_2,k_3)$ and the dimensionless shape function $\fnl^\text{(self2)}(k_1,k_2,k_3)$ as defined in \eqref{eq:fnl_defn} are given by
\begin{align}
\label{eq:3point_fun_R_fnl_gen} 
\left\langle \calR(\lk_1)\calR(\lk_2)\calR(\lk_3) \right\rangle = & (2\pi)^3 \delta^{(3)}\left(\sum_i \lk_i\right) B_\calR^\text{(self2)}(k_1,k_2,k_3)  \nonumber \\
= &  (2\pi)^3 \delta^{(3)}\left(\sum_i \lk_i\right) \left(\frac{3}{10} \right) \fnl^\text{(self2)}(k_1,k_2,k_3) \frac{k_1^3+k_2^3+k_3^3}{(k_1k_2k_3)^3} (2\pi)^4\calP_\calR^2 \, .
\end{align}
Here $H$ and $\dot{\phi}_0$ are evaluated at horizon exit. Factorising $\fnl^\text{(self2)}(k_1,k_2,k_3)$ in terms of the shape dependent and independent parts, we have
\begin{equation}
\fnl(k_1,k_2,k_3)^\text{(self2)} = \mathcal{F}^\text{(self2)} s^\text{(self2)}(k_1,k_2,k_3) \frac{(XYZ)^{3/2}}{X^3+Y^3+Z^3} \, .
\end{equation}
For convenience, we have introduced a normalisation wavenumber $K$ and have defined $x_i \equiv -K\tau_i$, $X\equiv k_1/K$, $X\equiv k_2/K$  and $Z\equiv k_3/K$ in the above expression. In the massless limit $\nu_\theta$, $\nu_\phi\approx 3/2$, $\mathcal{F}^\text{(self2)}$ is
\begin{equation}
\mathcal{F}^\text{(self2)}  =  \left(\frac{\dot{\phi}_0}{H}\right) \left(\frac{5\alpha^3\beta}{H^8r_0^6} \right) 2^{-3/2}\pi^{9/2} \, ,
\end{equation}
%
and the dimensionless shape function $s^\text{(self2)}(k_1,k_2,k_3)$ is 
\begin{align}
\label{eq:3point_fun_R_shape_fun_self}
&  s^\text{(self2)}(k_1,k_2,k_3)  \nonumber \\ 
\equiv & \Re\left\{  \int_{0}^{\infty} \frac{{\rm d}x_1}{x_1} \int_{x_1}^{\infty} \frac{{\rm d}x_2}{x_2} \int_{x_2}^{\infty} \frac{{\rm d}x_3}{x_3} \int_{x_3}^{\infty} \frac{{\rm d}x_4}{x_4} \,\, \Re\left[H_{\nu_\phi}^{(1)}(X x_1)\right] \nonumber \right. \\ 
\times &\left[ -\left(x_2\right)^{3/2} \Im \left[H_{\nu_\theta}^{(1)}(X x_1)H_{\nu_\theta}^{(2)}(X x_2)\right]
\Re \left[H_{\nu_\theta}^{(1)}\left(Z x_2\right)H_{\nu_\theta}^{(2)}\left(Z x_4\right)H_{\nu_\phi}^{(2)}\left(Z x_4\right)\right] \right. \nonumber \\
 \times & H_{\nu_\theta}^{(1)}\left(Y x_2\right)H_{\nu_\theta}^{(2)}\left(Y x_3\right)H_{\nu_\phi}^{(2)}\left(Y x_3\right)  \nonumber \\ 
+ & \left(x_3\right)^{3/2} \Re \left[H_{\nu_\phi}^{(1)}\left(Y x_2\right)\right]
\Im \left[H_{\nu_\theta}^{(2)}(X x_1)H_{\nu_\theta}^{(2)}\left(Y x_2\right) H_{\nu_\theta}^{(1)}(X x_3)H_{\nu_\theta}^{(1)}\left(Y x_3\right)\right] \nonumber \\ 
\times & H_{\nu_\theta}^{(1)}\left(Z x_3\right)H_{\nu_\theta}^{(2)}\left(Z x_4\right)H_{\nu_\phi}^{(2)}\left(Z x_4\right) \nonumber \\
- &  i \left(x_4\right)^{3/2}\Re \left[H_{\nu_\phi}^{(1)}\left(Y x_2\right)\right]\Re\left[H_{\nu_\phi}^{(1)}\left(Z x_3\right)\right] \nonumber \\
\times & \left.\left.  H_{\nu_\theta}^{(2)}(X x_1)H_{\nu_\theta}^{(2)}\left(Y x_2\right)H_{\nu_\theta}^{(2)}\left(Z x_3\right)
H_{\nu_\theta}^{(1)}(X x_4)H_{\nu_\theta}^{(1)}\left(Y x_4\right)H_{\nu_\theta}^{(1)}\left(Z x_4\right) \right] \right\}  + 5 \,\, {\rm perm.} 
\end{align}
Here $s^\text{(self2)}(k_1,k_2,k_3)$ diverges in the IR in the massless limit. Introducing a lower cut-off $x_c$ as before, one finds 
\begin{equation}
s^\text{(self2)}(k_1,k_2,k_3) \approx \left(\frac{44}{243}\right) \pi^{-9/2} \left(\log x_c\right)^4 \frac{X^3+Y^3+Z^3}{(XYZ)^{3/2}} \, ,
\end{equation}
to leading order. In the following, we briefly summarise how to compute the above result.

To compute the leading order term in the IR limit, we take the following asymptotic form of the Hankel function in the $x\ll 1$ limit 
\begin{align}
\label{eq:asymptotic_form_Hankel}
H_{3/2}^{(1)}(x) \rightarrow & -i \frac{2^{3/2}\Gamma(3/2)}{\pi} x^{-3/2} -  i \frac{2^{-1/2}\Gamma(3/2)}{\pi(1/2)} x^{1/2} \nonumber \\
&+ \left(- i \frac{\cos(3\pi/2)\Gamma(-3/2)}{2^{3/2}\pi} + \frac{1}{2^{3/2}\Gamma(5/2)}\right) x^{3/2} + ... 
\end{align}
whereas that of $H_{3/2}^{(2)}(x)$ is the complex conjugate of \eqref{eq:asymptotic_form_Hankel}. Here we can see the real and the imaginary parts scale as $\calO(x^{3/2})$ and $\calO(x^{-3/2})$ respectively at leading order. Using this, we can see how the integrand in \eqref{eq:3point_fun_R_shape_fun_self} scales with $x_c$. For the second line  in \eqref{eq:3point_fun_R_shape_fun_self}, at leading order in the IR limit, it scales as 
\begin{equation}
\label{eq:integrand_asympt_form_first_line}
\prod_{i=1}^{4} (x_i)^{-1} \Re \left[H_{\nu_\phi}^{(1)}(X x_1)\right] \rightarrow \prod_{i=1}^{4} (x_i)^{-1} \left[\frac{1}{2^{3/2}\Gamma(5/2)}\right](X x_1)^{3/2} + ...
\end{equation}
whereas for the third line, the leading order piece is
\begin{align}
\label{eq:integrand_asympt_form_second_line}
 & -\left(x_2\right)^{3/2} \Im \left[H_{\nu_\theta}^{(1)}(X x_1)H_{\nu_\theta}^{(2)}(X x_2)\right]
\Re \left[H_{\nu_\theta}^{(1)}\left(Z x_2\right)H_{\nu_\theta}^{(2)}\left(Z x_4\right)H_{\nu_\phi}^{(2)}\left(Z x_4\right)\right] \nonumber \\
\,\,\, &\rightarrow  \left[\frac{\Gamma(3/2)}{\Gamma(5/2)\pi}\right]^2 \frac{2^{5/2}\Gamma(3/2)}{\pi} \left(\frac{x_2}{x_1}\right)^{3/2} Z^{-3/2} + ...
\end{align}
Here we have taken $H_{\nu_\theta}^{(1)}(X x_1)$, $H_{\nu_\theta}^{(1)}\left(Z x_2\right)$ and either $H_{\nu_\theta}^{(2)}\left(Z x_4\right)$ or $H_{\nu_\phi}^{(2)}\left(Z x_4\right)$ to be imaginary. For the fourth line in \eqref{eq:3point_fun_R_shape_fun_self}, it must be real in order to keep the whole integrand real. The leading order term scales as
\begin{equation}
\label{eq:integrand_asympt_form_third_line}
 H_{\nu_\theta}^{(1)}\left(Y x_2\right)H_{\nu_\theta}^{(2)}\left(Y x_3\right)H_{\nu_\phi}^{(2)}\left(Y x_3\right) \rightarrow 2\left[\frac{\Gamma(3/2)}{\Gamma(5/2)\pi}\right] \left[\frac{2^{3/2}\Gamma(3/2)}{\pi}\right](Y x_2)^{-3/2} + ...
\end{equation}
Here we have taken $H_{\nu_\theta}^{(1)}\left(Y x_2\right)$ and either $H_{\nu_\theta}^{(2)}\left(Y x_3\right)$ or $H_{\nu_\phi}^{(2)}\left(Y x_3\right)$ to be imaginary. Together the above leading order pieces in the three lines therefore give the following integrand
\begin{equation}
\label{eq:integrand_asympt_form_A}
\prod_{i=1}^{4} (x_i)^{-1} \left[\frac{\Gamma(3/2)}{\Gamma(5/2)\pi}\right]^4\left[\frac{\Gamma(3/2)2^{7/2}}{\pi}\right] \left(\frac{X}{YZ}\right)^{3/2} \,,
\end{equation}
and the corresponding contribution to $s^\text{(self2)}(k_1,k_2,k_3)$ scales as $(\log x_c)^4$ in the IR limit. Taking other choices of the leading order pieces in the Hankel functions give subleading terms. For instance, taking $H_{\nu_\theta}^{(1)}(X x_1)$ to be real and $H_{\nu_\theta}^{(2)}(X x_2)$ to be imaginary instead in the second line, together with the above leading order terms in the first and third lines, the resulting integral scales at most as $(-\log x_c)^3$. 


Similarly, for the fifth line in \eqref{eq:3point_fun_R_shape_fun_self}, the leading order piece is 
\begin{align}
\label{eq:integrand_asympt_form_fourth_line}
 &  \left(x_3\right)^{3/2} \Re \left[H_{\nu_\phi}^{(1)}\left(Y x_2\right)\right]
\Im \left[H_{\nu_\theta}^{(2)}(X x_1)H_{\nu_\theta}^{(2)}\left(Y x_2\right) H_{\nu_\theta}^{(1)}(X x_3)H_{\nu_\theta}^{(1)}\left(Y x_3\right)\right] \nonumber \\
\,\,\, &\rightarrow (x_3)^{3/2} \left[\frac{\Gamma(3/2)}{\Gamma(5/2)\pi}\right]^2 \frac{2^{3/2}\Gamma(3/2)}{\pi} (X x_1)^{-3/2} \left[\left( \frac{X}{Y}\right)^{3/2} + \left(\frac{Y}{X}\right)^{3/2} \right]  + ...
\end{align}
Here we have taken $H_{\nu_\theta}^{(2)}(X x_1)$, $ H_{\nu_\theta}^{(2)}\left(Y x_2\right)$ and either $H_{\nu_\theta}^{(1)}(X x_3)$ or $H_{\nu_\theta}^{(1)}\left(Y x_3\right)$ to be imaginary. In order for the overall contribution together with the second and fifth lines to the integral to be real, the leading order piece in the sixth line must be real and is given by
\begin{equation}
\label{eq:integrand_asympt_form_fifth_line}
 H_{\nu_\theta}^{(1)}\left(Z x_3\right)H_{\nu_\theta}^{(2)}\left(Z x_4\right)H_{\nu_\phi}^{(2)}\left(Z x_4\right) \rightarrow \left[\frac{\Gamma(3/2)}{\Gamma(5/2)\pi}\right]\left[\frac{2^{5/2}\Gamma(3/2)}{\pi}\right](Z x_3)^{-3/2} + ...
\end{equation}
Here we have taken $H_{\nu_\theta}^{(1)}\left(Z x_3\right)$ and either $H_{\nu_\theta}^{(2)}\left(Z x_4\right)$ or $H_{\nu_\phi}^{(2)}\left(Z x_4\right)$ to be imaginary. Together the second, fifth and sixth lines in \eqref{eq:3point_fun_R_shape_fun_self} give the following leading order contribution to $s^\text{(self2)}(k_1,k_2,k_3)$ in the IR limit
\begin{equation}
\label{eq:integrand_asympt_form_B}
\prod_{i=1}^{4} (x_i)^{-1} \left[\frac{\Gamma(3/2)}{\Gamma(5/2)\pi}\right]^4\left[\frac{2^{5/2}\Gamma(3/2)}{\pi}\right] \left[\left(\frac{X}{YZ}\right)^{3/2} + \left(\frac{Y}{XZ}\right)^{3/2} \right]  \, .
\end{equation}
%
Again similarly, the leading order term in the seventh line in \eqref{eq:3point_fun_R_shape_fun_self} is 
\begin{equation}
\label{eq:integrand_asympt_form_sixth_line}
- i\left(x_4\right)^{3/2} \Re \left[H_{\nu_\phi}^{(1)}\left(Y x_2\right)\right]\Re \left[H_{\nu_\phi}^{(1)}\left(Z x_3\right)\right] \rightarrow - i(x_4)^{3/2} \left[\frac{1}{2^{3/2}\Gamma(5/2)}\right]^2 (Y x_2)^{3/2} (Z x_3)^{3/2} + ...
\end{equation}
whereas for the eighth line, the leading order term must be imaginary in order to keep the overall contribution real and thus is given by
\begin{align}
\label{eq:integrand_asympt_form_seventh_line}
 & H_{\nu_\theta}^{(2)}(X x_1)H_{\nu_\theta}^{(2)}\left(Y x_2\right)H_{\nu_\theta}^{(2)}\left(Z x_3\right)
H_{\nu_\theta}^{(1)}(X x_4)H_{\nu_\theta}^{(1)}\left(Y x_4\right)H_{\nu_\theta}^{(1)}\left(Z x_4\right)  \nonumber \\
& \rightarrow i \left[\frac{2^{3/2}\Gamma(3/2)}{\pi}\right]^4 \left[\frac{\Gamma(3/2)}{\Gamma(5/2)\pi}\right] (XYZ)^{-3/2} ( x_1x_2x_3)^{-3/2} \left[\left( \frac{X}{YZ}\right)^{3/2} +  \left( \frac{Y}{XZ}\right)^{3/2} + \left( \frac{Z}{XY}\right)^{3/2} \right]  + ...
\end{align}
Here we have taken either $H_{\nu_\theta}^{(1)}(X x_4)$, $H_{\nu_\theta}^{(1)}\left(Y x_4\right)$ or $H_{\nu_\theta}^{(1)}\left(Z x_4\right)$ to be real. Together the second, seventh and eighth lines give the following leading order contribution to $s^\text{(self2)}(k_1,k_2,k_3)$ in the IR limit
\begin{equation}
\label{eq:integrand_asympt_form_C}
\prod_{i=1}^{4} (x_i)^{-1} \left[\frac{\Gamma(3/2)}{\Gamma(5/2)\pi}\right]^4 \frac{2^{3/2}\Gamma(3/2)}{\pi} \left[\left( \frac{X}{YZ}\right)^{3/2} +  \left( \frac{Y}{XZ}\right)^{3/2} + \left( \frac{Z}{XY}\right)^{3/2} \right]   \, . 
\end{equation}
After factorising out the common factor $\prod_{i=1}^{4} (x_i)^{-1}$ in \eqref{eq:integrand_asympt_form_A}, \eqref{eq:integrand_asympt_form_B} and \eqref{eq:integrand_asympt_form_C}, we can see the integral goes as
\begin{equation}
\label{eq:integral_IR_limit_scaling}
\int_{x_c} {\rm d}x_1 \int_{x_1} {\rm d}x_2 \int_{x_2} {\rm d}x_3 \int_{x_3} {\rm d}x_4 \prod_{j=1}^{4} (x_i)^{-1} \sim \frac{(\log x_c)^4}{4!} \,,
\end{equation}
assuming the integral vanishes in the UV limit $x_i\rightarrow \infty$. Thus finally, to leading order in the IR limit, $s^\text{(self2)}(k_1,k_2,k_3)$ is given by
\begin{align}
\label{eq:s_self_IR_limit}
s^\text{(self2)}(k_1,k_2,k_3) \underset{\rm IR}{\rightarrow} & (XYZ)^{-3/2}\left[\frac{\Gamma(3/2)}{\Gamma(5/2)\pi}\right]^4\left[\frac{2^{3/2}\Gamma(3/2)}{\pi}\right] \frac{(\log x_c)^4}{4!}\nonumber \\
& \times \left[4X^3 + 2(X^3+Y^3) + (X^3 + Y^3 + Z^3) \right] + \,\,5 \,\, {\rm perm.} \, .
\end{align}
Summing over all permutations, it is not difficult to see $s^\text{(self2)}(k_1,k_2,k_3)$ scales as $(X^3+Y^3+Z^3)/(XYZ)^{3/2}$ and thus $\fnl(k_1,k_2,k_3)^\text{(self2)}$ is approximately shape independent. Using the results of the cross-correlation spectrum $\calP_\calC$ and expressing $\fnl(k_1,k_2,k_3)^\text{(self2)}$ in terms of $\calP_\calC$, we therefore arrive at \eqref{fnls_1}. 

\subsubsection*{C2. Contribution from $H_3^\text{(cross2)}$}

The corresponding leading order contribution to the inflaton bispectrum $\left\langle \phi^3 \right\rangle$ from the axion cross-interaction $H_3^\text{(cross2)}$ is of quadratic order in $H^I_2$. Expressed in commutator form, this contribution is given by
\begin{equation}
\label{eq:3point_fun_inflaton_cross}
\left\langle \phi^3 \right\rangle^\text{(cross2)} = \int_{t_0}^{t} {\rm d}t_1 \int_{t_0}^{t_1} {\rm d}t_2 \int_{t_0}^{t_2} {\rm d}t_3 \left\langle\left[H^I(t_3),\left[H^I(t_2),\left[H^I(t_1),\phi_I(t)^3 \right]\right]\right]\right\rangle \,,
\end{equation}
where one of the $H^I$ is given by $H_3^\text{(cross2)}$. Expanding this, we get
\begin{align}
\label{eq:3point_fun_inflaton_B_expand}
\left\langle \phi^3 \right\rangle^\text{(cross2)} =& -4\alpha^2\gamma \left(\prod_{i=1}^3 u_{k_i}(0)\right)(2\pi)^3\delta^{(3)}\left(\lk_1+\lk_2+\lk_3\right) \nonumber \\
\times & \Re \left[-i\int_{-\infty}^{0}{\rm d}\tau_1\int_{-\infty}^{\tau_1}{\rm d}\tau_2\int_{-\infty}^{\tau_2}{\rm d}\tau_3 \prod_{i=1}^3 a^4(\tau_i)\left(D+E+F\right)\right] + 5\,\, {\rm perm.} \,,
\end{align}
where $D$, $E$ and $F$ correspond to contributions coming from replacing $H_I(t_1)$, $H_I(t_2)$ and $H_I(t_3)$ respectively with $H_I^{(3),{\rm cross}}$:
\begin{align}
\label{eq:3point_fun_inflaton_B_terms}
& D =  [u_{k_1}(\tau_1) - {\rm {\rm c.c.}}][v_{k_2}^*(\tau_1)v_{k_2}(\tau_2)u_{k_2}(\tau_2) - {\rm c.c.}] v_{k_3}(\tau_1)v_{k_3}^*(\tau_3)u_{k_3}^*(\tau_3) \, , \\
& E =  [u_{k_1}(\tau_1) - {\rm c.c.}][v_{k_1}^*(\tau_1)v_{k_1}(\tau_2)u_{k_2}(\tau_2) - {\rm c.c.}] v_{k_3}(\tau_2)v_{k_3}^*(\tau_3)u_{k_3}^*(\tau_3)  \, ,\\
& F = [u_{k_1}(\tau_1) - {\rm c.c.}][u_{k_2}(\tau_2) - {\rm c.c.}]v_{k_1}(\tau_1)v_{k_2}(\tau_2)v_{k_1}^*(\tau_3)v_{k_2}^*(\tau_3)u_{k_3}^*(\tau_3) \, ,
\end{align}
Here we have again used the fact that $u_{k}(0)$ is real. Written in terms of the Hankel functions, they are given by
\begin{align}
\label{eq:3point_fun_inflaton_B_terms}
D = & i \calN (-\tau_1)^{9/2}(-\tau_2)^{3}(-\tau_3)^{3} \Re \left[H_{\nu_\phi}^{(1)}(-k_1\tau_1)\right] \Re\left[H_{\nu_\theta}^{(2)}(-k_2\tau_1)H_{\nu_\theta}^{(1)}(-k_2\tau_2)H_{\nu_\phi}^{(1)}(-k_2\tau_2)\right] \nonumber \\
\times & H_{\nu_\theta}^{(1)}(-k_3\tau_1)H_{\nu_\theta}^{(2)}(-k_3\tau_3)H_{\nu_\phi}^{(2)}(-k_3\tau_3) \, ,\\
E = & - i \calN (-\tau_1)^{3}(-\tau_2)^{9/2}(-\tau_3)^{3} \Re \left[H_{\nu_\phi}^{(1)}(-k_1\tau_1)\right]\Re \left[H_{\nu_\theta}^{(2)}(-k_1\tau_1)H_{\nu_\theta}^{(1)}(-k_1\tau_2)H_{\nu_\phi}^{(1)}(-k_2\tau_2)\right] \nonumber \\
\times & H_{\nu_\theta}^{(2)}(-k_3\tau_2) H_{\nu_\theta}^{(1)}(-k_3\tau_3)H_{\nu_\phi}^{(1)}(-k_3\tau_3) \, ,\\
F = & -i \calN (-\tau_1)^{3}(-\tau_2)^{3}(-\tau_3)^{9/2} \Re\left[H_{\nu_\phi}^{(1)}(-k_1\tau_1)\right] \Re\left[H_{\nu_\phi}^{(1)}(-k_2\tau_2)\right]  \nonumber \\
\times & H_{\nu_\theta}^{(2)}(-k_1\tau_1)H_{\nu_\theta}^{(2)}(-k_2\tau_2)
 H_{\nu_\theta}^{(1)}(-k_1\tau_3)H_{\nu_\theta}^{(1)}(-k_2\tau_3)H_{\nu_\phi}^{(1)}(-k_3\tau_3) \, ,
\end{align}
where $\calN\equiv 4/r_0^4(H\sqrt{\pi}/2)^7$. The resulting the curvature bispectrum $B_\calR^\text{(cross2)}(k_1,k_2,k_3)$ is simply given by multiplying \eqref{eq:3point_fun_inflaton_B_expand} with $-(H/\dot{\phi}_0)^3$. Similar to the $H_3^\text{(self2)}$ case, we factorise the corresponding shape function $\fnl^\text{(cross2)}(k_1,k_2,k_3)$ into shape independent and dependent parts as follows
\begin{equation}
\fnl(k_1,k_2,k_3)^\text{(cross2)} = \mathcal{F}^\text{(cross2)} s^\text{(cross2)}(k_1,k_2,k_3) \frac{(XYZ)^{3/2}}{X^3+Y^3+Z^3} \, .
\end{equation}
In the massless limit $\nu_\theta$, $\nu_\phi\approx 3/2$, $\mathcal{F}^\text{(cross2)}$ is given by
\begin{equation}
\mathcal{F}^\text{(cross2)}  =  \left(\frac{\dot{\phi}_0}{H}\right) \left(\frac{\alpha^2\gamma}{H^6r_0^4} \right) \left(\frac{\pi}{2}\right)^{7/2} \, ,
\end{equation}
whereas $s^\text{(cross2)}(k_1,k_2,k_3)$ is 
\begin{align}
\label{eq:3point_fun_R_shape_fun_cross}
 & s^\text{(cross2)}(k_1,k_2,k_3)  \nonumber \\
 = & \Re\left\{\int_{0}^{\infty} \frac{{\rm d}x_1}{x_1} \int_{x_1}^{\infty} \frac{{\rm d}x_2}{x_2} \int_{x_2}^{\infty} \frac{{\rm d}x_3}{x_3}  \,\, \Re \left[H_{\nu_\phi}^{(1)}(X x_1)\right] 
\left\{\left(x_1\right)^{3/2} \Re \left[H_{\nu_\theta}^{(2)}\left(Y x_1\right) H_{\nu_\theta}^{(1)}\left(Y x_2\right)H_{\nu_\phi}^{(1)}\left(Y x_2\right)
\right] \right.\right.  \nonumber \\
\times & H_{\nu_\theta}^{(1)}\left(Z x_1\right)H_{\nu_\theta}^{(2)}\left(Z x_3\right) H_{\nu_\phi}^{(2)}\left(Z x_3\right)  
- \left(x_2\right)^{3/2} \Re\left[H_{\nu_\theta}^{(2)}(X x_1) H_{\nu_\theta}^{(1)}(X x_2)H_{\nu_\phi}^{(1)}\left(Y x_2\right)
\right] \nonumber \\ 
\times & H_{\nu_\theta}^{(2)}\left(Z x_2\right) H_{\nu_\theta}^{(1)}\left(Z x_3\right)H_{\nu_\phi}^{(1)}\left(Z x_3\right) 
- \left(x_3\right)^{3/2} \Re\left[H_{\nu_\phi}^{(1)}\left(Y x_2\right)\right]  H_{\nu_\theta}^{(2)}(X x_1)H_{\nu_\theta}^{(2)}\left(Y x_2\right) \nonumber \\
\times & \left. \left. H_{\nu_\theta}^{(1)}(X x_3)
H_{\nu_\theta}^{(1)}\left(Y x_3\right)H_{\nu_\phi}^{(1)}\left(Z x_3\right)\right\} \right\} 
+   5 \,\, {\rm perm.} 
\end{align}
Here we have again defined $x_i \equiv -K\tau_i$, $X\equiv k_1/K$, $Y\equiv k_2/K$ and $Z\equiv k_3/K$. Similar to the $H_3^\text{(self2)}$ case, $s^\text{(cross2)}(k_1,k_2,k_3)$ diverges in the IR in the massless limit. Introducing a lower cut-off $x_c$ and following similar approach as in the $H_3^\text{(self2)}$ case, one finds  
\begin{align}
\label{eq:s_cross_IR_limit}
 s^\text{(cross2)}(k_1,k_2,k_3) \underset{\rm IR}{\rightarrow} & -(XYZ)^{-3/2}\left[\frac{\Gamma(3/2)}{\Gamma(5/2)\pi}\right]^3\left[\frac{2^{3/2}\Gamma(3/2)}{\pi}\right] \left[\frac{(\log x_c)^{3}}{3!}\right]\nonumber \\
&  \left[4X^3 - 2(X^3+Y^3) - (X^3 + Y^3+Z^3) \right] + \,\,5 \,\, {\rm perm.} \, .
\end{align}
Summing over all permutations, it is not difficult to see $s^\text{(cross2)}(k_1,k_2,k_3)$ also scales as $(X^3+Y^3+Z^3)/(XYZ)^{3/2}$ and thus $\fnl(k_1,k_2,k_3)^\text{(cross2)}$ is approximately shape independent. Rewriting $\fnl(k_1,k_2,k_3)^\text{(cross2)}$ in terms of $\calP_\calC$, we therefore arrive at \eqref{fnls_1}.

\end{document}